# Response to "Comment on 'Phase transition temperatures of 405-725 K in superfluid ultra-dense hydrogen clusters on metal surfaces' [AIP Advances 6, 045111 (2016)]".

Leif Holmlid
Atmospheric Science, Department of Chemistry and Molecular Biology, University of Gothenburg, SE-412 96 Göteborg, Sweden. Tel.:+46-738397830. E-mail address: holmlid@chem.gu.se

Bernhard Kotzias
Kotzias Hydrogen Energy Consulting, Bremen, Germany. Tel.; +49-1794291715. E-mail address: bernhard.kotzias@freenet.de.

## Abstract

In this invited response we answer all comments by Engelen and Hansen. We point out that the superfluid and superconductive properties of H(0) have been published previously. We explain some differences between covalently bonded molecules and the molecules in the ultradense matter H(0) form, and explain some aspects of the energetics of H(0) molecules during Coulomb explosions. We point out that the experimental spectra shown in our publication are not ion time-of-flight spectra but neutral time-of-flight spectra with the peak width given by the internal energetics and not by experimental factors. We point out that no phase diagram has been measured for H(0). Further we point out that a Rydberg state is a hydrogenic state and thus that all hydrogen atom states are Rydberg states. That Rydberg states always have large principal quantum numbers is a complete misunderstanding. We point out that a QM description of H(0) is published. We point out that the internuclear distances in p(0), D(0) and pD(0) have been measured by rotational spectroscopy in two publications for three different spin states. They are measured in the pm range with fm precision.

We do not share the view by the authors of this comment that new results in physics are an offence to all scientists who have worked with similar studies previously. On the contrary, new results build on the previous results and may even replace them, that is how science works. Physics is not a set of eternal laws but a dynamical description of the world we have around us now.

The reason for our experimental study of the transition temperatures from a superfluid to a normal state was that results on the superfluid and superconductive state of H(0) had been published by one of us. If such a phase exists, there must also exist a transition temperature and the goal of our study was to observe and measure the transition temperatures, which we did.

We do not repeat the conclusions of our paper further, even if it seems that the authors of the comment have not understood them, That may of course be due to many unknown factors. The main subject of our paper, the phase transition temperatures are apparently accepted by the authors of the comment since they have no scientific remarks on this central point. We include a small scientific comment on the transition temperatures at the end of this response. We only answer the scientific parts of the comments. We do not think it is worthwhile to react to all comments about what is "established facts in several fields" or not. We only want to remark that if H(0) had been studied earlier the "established facts in several fields" would be different.

<u>We cite statements from their comment and answer point-by-point.</u>

"Already the title of the paper announces a couple of truly revolutionary discoveries, if they had been correct. Superfluidity has to date been…"

It is clear that the authors of the comment have missed the papers where these discoveries are published, refs. 5, 6 and 7 in our paper.

"this positive energy should be compensated by at least the same amount of negative energy supplied by the electrons. Otherwise the compound would not be stable". This statement shows that the authors do not understand how a chemical bond of any type is formed. The important point is the *interaction* between charges of different sign which is attractive. This gives a bonding negative potential energy.

"The authors' assignment requires that the electrons absorb at least 640 eV in order to produce the charged particles." From this statement it is again clear that the authors of the comment do not understand what a chemical bond is. The electrons do not absorb 640 eV: how can an electron absorb energy? The authors do not recognize that it is the *interaction* between the electrons and nuclei that gives the bonding. Instead of giving a basic description of chemical bonding here we have to refer to the discussion in the H(0) review paper published in 2019 http://doi.org/10.1088/1402-4896/ab1276. Figure 1 in that paper should provide sufficient information for the authors to reach a better understanding.

"The authors fail to explain how an energy of this magnitude is imparted to the molecule." The molecule the authors talk about is $H_{2N}$ with N up to at least 30. The total bond energy is of the order of 10 keV. 640 eV is a negligible part of the energy available in the H(0) molecule. Besides, this energy does not have to be imparted to the molecule at all. See our answers above.

"In fact, they even explicitly state that the electrons are easily removed by the laser pulse, and seemingly ignore the question entirely." No, we do not ignore this question. However, that the electrons are easily removed agrees with H(0) being a super material. We do not understand all aspects of H(0) neither 2016 or now. There are understandably no previous studies of the electron energies for similar materials from the great days of superfluids. However, we find agreement with the theory for superconductors by Hirsch in these aspects. In 2016 when our paper in AIP Advances was published, we were not aware of the description of matter given by Hirsch on three different length and energy scales, also described in the 2019 H(0) review. Thus we could not write anything about this which might have been helpful for the serious reader.

"The experimental results presented by the authors consist of four time-of-flight spectra. They are all of rather poor resolution, with large amounts of unresolved intensity at late times." From this comment it is clear that the authors believe the spectra to be time-of-flight mass spectrometry spectra where the resolution is influenced by experimental parameters. This is not so since the spectra are pure neutral time-of-flight spectra with the width of the peaks given by the energy release from the Coulomb explosions. These widths are thus inherent in the energetics of H(0).

"In our experience the spectra rather look like the manifestation of a charging effect in the equipment, although other instrumental artifacts may also contribute." The experience of the authors is clearly insufficient since the time-of flight spectra are due to neutral particles which are quite difficult to influence by any charging effects. The spectra are not due to ions. In our paper we clearly write "The fast particles impact on a steel catcher foil in the detector". Ions are not mentioned. A catcher foil is of course only used for fast neutral particles, not for ions. The authors of the comment did not notice this very important point.

"For example, the hydrogen molecule is known from spectroscopy to have an equilibrium length of 0.74 Å [4]. Any reduction of this distance to the proposed 2 picometer grossly contradicts all this knowledge." H(0) does not contain covalently bonded hydrogen molecules. There is no contradiction.

"This should even be clear from an understanding of quantum mechanics at an elementary level". The description of H(0) in QM terms is given in the 2019 review, maybe not elementary.

"Hydrogen has been liquified for decades and its phase diagram is well known. Nowhere does a phase with the claimed extremely high density appear." One needs to be much more precise. A gas of covalently bonded hydrogen molecules has been liquefied and the phase diagram for such a phase is well known. H(0) does not consist of covalently bonded molecules but consists of hydrogen atoms with electron angular momentum equal to zero bound in molecules mainly of the form $H_{2N}$. No attempt has ever been made to measure the phase diagram of such a superfluid phase. There have been some work made on including ordinary Rydberg Matter in a phase diagram, but even that kind of simpler system apparently contains other dimensions not easily representable in a phase diagram.

"The word Rydberg implies that atoms have been excited to high-lying states, with their concomitant large principal quantum numbers and sizes. Principal quantum numbers of at least four or five are involved." One needs to be much more precise. A Rydberg state follows the Rydberg formula (please consult Wikipedia for exact information, do not consult textbooks in physics) which means that all states of the hydrogen atom are Rydberg states also called hydrogenic states consisting of one electron around a nucleus. For many-electron atoms, the easiest way to make a Rydberg state is to excite an electron to a high *n* value so the core ion mimicks a nuclei. Such a state is just a special case of a Rydberg state.

"The known large physical dimensions of Rydberg states are now postulated by the authors, without any argument, to be converted into extremely small values, more than an order of magnitude smaller than the Bohr radius." This is a continuation of the misunderstanding about Rydberg states, see above. Besides, the size of any atomic state is not important since it is the internuclear distance in the molecule which is important. The electronic state in the molecule is not the same as the atomic state forming it. The internuclear distances in H(0) are measured with three main methods. See the review from 2019. The simplest method to understand of these three is probably the rotational spectroscopy with results given in two publications: 10.1016/j.molstruc.2016.10.091 and 10.1016/j.molstruc.2018.06.116 and in the review paper from 2019. The interatomic distances are measured in p(0), D(0) and pD(0) to be in the pm range for three spin states with a precision in the few fm range. So the dimensions are not postulated but measured. We could also give theoretical arguments for the small internuclear distances but this has already been published in the 2019 review following the description by Hirsch.

"The claims are based on evidence which with the best of wills can only be considered very flimsy, and can only be made with a complete disregard of well established scientific facts." There is no point in our paper where the scientific facts are altered due to this comment. So we think they are well established.

The results on H(0) have been published after peer review in around 65 scientific papers. Thus we feel that the authors of this comment to our paper are in fact attacking the well established system with peer review for scientific publishing. They apparently want to replace peer review with a superior editor who knows all fields of research and can take the decisions to reject or publish without advice. As can be seen here from the performance of these self-elected reviewers of our six year old paper such a superior judge is hardly possible. They selected our paper as suitable for their purposes and could have prepared for this self-chosen task but did not.

If we finally return to science instead of education we want to add a basic point about superfluidity. This is the subject of our paper, but there are few comments on this by the authors. We cannot yet provide a microscopic model of ultra-dense hydrogen, but can remind about the well- known formula for the transition temperature of Bose-Einstein-Condensates, which is $T \sim h^2/m \, n^{2/3}$. Since the density of H(0) is a factor >1000 higher than for ordinary matter and the mass of the superfluid particles is low, 1 or 2 u, the transition temperature may well be 100 times higher than for other superfluids, in agreement with our experimental results. If this theoretical description is fully applicable to H(0) is another question.